\documentclass[amsmath,amssymb]{revtex4}

\usepackage{amsmath}
\usepackage{graphicx}
\usepackage{dcolumn}
\usepackage{bm}
\usepackage{hyperref}
\usepackage{orcidlink}

\newcommand{\Tr}{{\rm Tr}}
\newcommand{\myslash}[1]{#1\!\!\!/}
\begin{document}


\title{$P_c$ states and their open-charm decays with the complex scaling method}

\author{Zi-Yang Lin\orcidlink{0000-0001-7887-9391}}
 \email{lzy\_15@pku.edu.cn}
\author{Jian-Bo Cheng\orcidlink{0000-0003-1202-4344}}%
 \email{jbcheng@pku.edu.cn}
\author{Bo-Lin Huang\orcidlink{0000-0003-1202-4475}}
 \email{blhuang@pku.edu.cn}

\author{Shi-Lin Zhu\orcidlink{0000-0002-4055-6906}}
 \email{zhusl@pku.edu.cn}

\affiliation{
  School of Physics and Center of High Energy Physics, Peking University 10087, China
}%

\date{\today}%

\begin{abstract}
A partial width formula is proposed using the analytical extension
of the wave function in momentum space. The distinction of the
Riemann sheets is explained from the perspective of the Schrodinger
equation. The analytical form in coordinate space and the partial
width are derived subsequently. Then a coupled-channel analysis is
performed to investigate the open-charm branching ratios of the
$P_c$ states, involving the contact interactions and
one-pion-exchange potential with the three-body effects. The low energy constants
are fitted using the experimental masses and widths as input. The
$P_c(4312)$ is found to decay mainly to $\Lambda_c\bar{D}^*$, while
the branching ratios of the $P_c(4440)$ and $P_c(4457)$ in different
channels are comparable. Under the reasonable assumption that the
off-diagonal contact interactions are small, the $J^P$ quantum
numbers of the $P_c(4440)$ and the $P_c(4457)$ prefer $\frac{1}{2}^-$
and $\frac{3}{2}^-$ respectively. Three additional $P_c$ states at
4380 MeV, 4504 MeV and 4516 MeV, together with their branching
ratios, are predicted. A deduction of the revised one-pion-exchange
potential involving the on-shell three-body intermediate states is
performed.
\end{abstract}

\maketitle

\section{Introduction}
Searching for exotic states composed of four or more quarks has been
a hot topic in hadron physics
\cite{Chen:2016qju,Lebed:2016hpi,Olsen:2017bmm,Guo:2017jvc,Liu:2019zoy,Brambilla:2019esw,Chen:2022asf,Meng:2022ozq}.
Since the discovery of the hidden-charm pentaquarks, the $P_c$
states, (also named as $P_{\psi}^N$ in line with the new naming
convention for the exotic states \cite{Gershon:2022xnn}), have been
investigated in a variety of works. Although the $P_c$ states are
commonly believed to be the bound states of the charmed mesons and
baryons, only their masses or binding energies can be obtained from
the Schr\"odinger equation. In this work, we aim to derive their
branching ratios in the open-charm channels, together with their
masses and total widths in a self-consist framework. We apply the
complex scaling method (CSM) to a coupled-channel analysis and
explain how the analytical extension of the wave function works.
Besides, we include the effect of the on-shell three-body
intermediate states.

The $P_c(4380)$ and $P_c(4450)$ states were first observed in the
$J/\psi p$ invariant mass spectrum in the $\Lambda_b^0\rightarrow
J/\psi pK^-$ decays by the LHCb Collaboration in 2015
\cite{LHCb:2015yax,LHCb:2016ztz}. They carried out a more precise
analysis with a larger data sample in 2019, and discovered a new
state $P_c(4312)$ and the two peak structure of the $P_c(4450)$,
namely $P_c(4440)$ and $P_c(4450)$ \cite{LHCb:2019kea}. Their masses
and widths are listed in Table \ref{tab:exp}, which are fitted under
the incoherent relativistic Breit-Wigner assumptions. Additionally,
the evidence of a new structure $P_c(4337)$ was found in the
$B_s^0\rightarrow J/\psi p\bar{p}$ decays in 2021
\cite{LHCb:2021chn}.

\begin{table}
  \renewcommand\arraystretch{1.5}
  \caption{\label{tab:exp}The $P_c$ states reported in Ref.~\cite{LHCb:2019kea}.}
  \begin{ruledtabular}
  \begin{tabular}{ccc}
    State&$M$[MeV]&$\Gamma$[MeV]\\
    \hline
    $P_c(4312)^+$ & $4311.9\pm 0.7^{+6.8}_{-0.6}$& $9.8\pm2.7^{+3.7}_{-4.5}$\\
    $P_c(4440)^+$ & $4440.3\pm 1.3^{+4.1}_{-4.7}$& $20.6\pm4.9^{+8.7}_{-10.1}$\\
    $P_c(4457)^+$ & $4457.3\pm 0.6^{+4.1}_{-1.7}$& $6.4\pm2.0^{+5.7}_{-1.9}$
  \end{tabular}
  \end{ruledtabular}
\end{table}

These pentaquarks lie close to and below the
$\Sigma_c^{(*)}\bar{D}^{(*)}$ thresholds, and are commonly believed
to be the hadronic molecules, which were first predicted in Refs.
\cite{Yang:2011wz,Wu:2010jy,Wu:2010vk}. Since their discovery, they have been investigated in the
frameworks of the quark model \cite{Hiyama:2018ukv,Weng:2019ynv},
kinetical effects \cite{Guo:2015umn,Bayar:2016ftu}, compact states
coupled to the meson-hadron channels \cite{Yamaguchi:2019seo}, the
QCD sum rule \cite{Chen:2019bip}, the vector-meson-exchange model
\cite{Wu:2012md,Dong:2021bvy}, the one-boson-exchange model \cite{Chen:2019asm},
the large $N_c$ approximation \cite{Eides:2015dtr} and the chiral
effective field theory \cite{Meng:2019ilv,Wang:2019ato}. (For a
detailed review, see
Ref.~\cite{Liu:2019zoy,Brambilla:2019esw,Meng:2022ozq}).

Since the pion exchanges provide a strong coupling between the
$\Sigma_c^{(*)}\bar{D}^{(*)}$ channels, a coupled-channel analysis
is required in the molecule picture. The $P_c$ states are most
likely to be the quasi-bound states (Feshbach resonances). They are
on the first Riemann sheet with respect to the higher thresholds and
on the second Riemann sheet with respect to the lower thresholds. In
a coupled-channel analysis, the two-body open-charm decay widths
will be obtained automatically. In
Refs.~\cite{Xiao:2019aya,Feijoo:2022rxf,Du:2019pij,Burns:2022uiv},
the widths and $T$ matrix residues of the $P_c$ states are obtained
through the Bethe-Salpeter equation or the Lippmann-Schwinger
equation (LS equation).

The three-body effect should be taken into account since the mass
difference between the $D(\Lambda_c)$ and $D^*(\Sigma_c)$ is
comparable to the pion mass. The large transferred energy of the
exchanged pion results in a retarded potential in coordinate space.
Besides, the intermediate $\Lambda_c \bar{D}\pi$ state can be
on-shell and contribute to the total width. In
Ref.~\cite{Du:2023hlu}, the authors discussed the three-body effect
and the consequent left-hand or right-hand cut in the $T_{cc}^+$
state. When the mass difference is larger than the pion mass, the
one-pion-exchange (OPE) potential introduces a right-hand cut at the
three-body threshold and the quasi-bound state lies on the second
Riemann sheet with respect to the three-body threshold. In
Ref.~\cite{Du:2019pij,Du:2021fmf}, the authors calculated the pole
positions and couplings of the $P_c$ states in the framework of the
time-ordered-perturbation theory and the LS equation. In this work,
we retain both the relativistic form and the three-body effect in
the OPE potential and adopt the complex scaling method (CSM).

The CSM is a shortcut to derive the bound states and resonances
simultaneously by extending the Schrodinger equation to the complex
plane \cite{CSM-Moiseyev,Myo:2014ypa}. In Ref.~\cite{Lin:2022wmj},
we put forward the CSM in momentum space to investigate the
$T_{cc}^+$ and $X(3872)$. In this work, we further study the complex
wave function and develop the partial width formula. Moreover, we
systematically illustrate the CSM in momentum space from the point
of view of analytical extension. In principle, with an appropriate
choice of the integral path, we can solve the poles of the $T$
matrix on any Riemann sheets. Similar transformations can be
employed in LS equations \cite{Chen:2022wkh}.

This paper is organized as follows. In Sec.~\ref{sec:Lagrangian}, we
introduce the Lagrangians including the OPE and contact terms. In
Sec.~\ref{sec:CSM} and \ref{sec:partialwidth}, we explain the CSM in
momentum space and its application to calculate the partial width.
In Sec.~\ref{chap:3bodies}, we derive the OPE potential involving
the three-body effect. Then in Sec.~\ref{sec:Veff}, we present the
parameters and effective potentials. In Sec.~\ref{sec:numeric}, we
fit the low energy constants (LEC) using the masses and widths of
the $P_c(4312)$, $P_c(4440)$ and $P_c(4457)$ as inputs and predict
their branching ratios in the open-charm channels. We predict
several states. Sec.~\ref{sec:sum} is a summary.

\section{Lagrangian\label{sec:Lagrangian}}

We interpret the $P_c(4312)$, $P_c(4440)$ and $P_c(4457)$ as the
$\Sigma_c\bar{D}^{(*)}$ molecules. A coupled-channel analysis is
performed in the $\Sigma_c^{(*)}\bar{D}^{(*)}$ system. The
Lagrangian is constructed under heavy quark spin symmetry (HQSS)
\cite{Wise:1992hn}. The leading order contact terms and OPE are
included.

Following Refs.~\cite{Wang:2019ato,Chen:2022iil}, we organize the
$\Lambda_c$ and $\Sigma_c^{(*)}$ states into an iso-singlet and an
iso-triplet. The matrices in $SU(2)$ flavor space read
\begin{eqnarray}
  \psi_{1}=\left[\begin{array}{cc}
      0 & \Lambda_{c}^{+} \\
      -\Lambda_{c}^{+} & 0
      \end{array}\right], \quad
      \psi_{3}=\left[\begin{array}{cc}
      \Sigma_{c}^{++} & \frac{\Sigma_{c}^{+}}{\sqrt{2}} \\
      \frac{\Sigma_{c}^{+}}{\sqrt{2}} & \Sigma_{c}^{0}
      \end{array}\right],\quad
      \psi_{3^{*}}^{\mu}=\left[\begin{array}{cc}
      \Sigma_{c}^{*++} & \frac{\Sigma_{c}^{*+}}{\sqrt{2}} \\
      \frac{\Sigma_{c}^{*+}}{\sqrt{2}} & \Sigma_{c}^{* 0}
      \end{array}\right]^{\mu},
\end{eqnarray}
where "1" denotes the iso-singlet, and "3" denote the iso-triplet.
The $\psi_{3^{*}}^{\mu}$ denotes the spin-$\frac{3}{2}$
Rarita-Schwinger field.

The $\Sigma_c$ and $\Sigma_c^*$ form a multiplet under HQSS, which
can be arranged into a superfield,
\begin{eqnarray}
   \psi^{\mu}=\mathcal{B}_{3^{*}}^{\mu}-\frac{1}{\sqrt{3}}\left(\gamma^{\mu}+v^{\mu}\right) \gamma^{5} \mathcal{B}_{3},\quad \bar{\psi}^\mu=\bar{\mathcal{B}}_{3^{*}}^{\mu}+\frac{1}{\sqrt{3}}\bar{\mathcal{B}}_{3}\gamma^{5}\left(\gamma^{\mu}+v^{\mu}\right),
\end{eqnarray}
where $\mathcal{B}_{i}$ ($i=1,3,3^*$) is the light components of the
heavy baryon fields,
\begin{eqnarray}
  \mathcal{B}_{i}=e^{i M_{i} v \cdot x} \frac{1+\myslash{v}}{2} \psi_{i},\qquad \mathcal{H}_{i}=e^{i M_{i} v \cdot x} \frac{1-\myslash{v}}{2} \psi_{i}.
\end{eqnarray}
Under the heavy quark symmetry, only the light components with the
projection operator $\frac{1+\myslash{v}}{2}$ survive in the leading
order. The heavy components only contribute to $1/M_Q$ corrections
and vanish when the heavy quark mass $M_Q\rightarrow\infty$.
Considering the $SU(2)$ chiral symmetry in flavor space, the LO
Lagrangian reads,
\begin{eqnarray}
  \mathcal{L}_{\mathcal{B}\phi}=-\Tr(\bar{\psi}^\mu iv\cdot D\psi^\mu)+\frac{i\delta_a}{2}\Tr(\bar{\psi}^\mu\sigma_{\mu\nu}\psi^\nu)+ig_1\epsilon_{\mu\nu\rho\sigma}\Tr\left(\bar{\psi}^\mu u^\rho v^\sigma \psi_\nu\right)+g_2\Tr\left(\bar{\psi}^\mu u_\mu \mathcal{B}_1+\text{h.c.}\right),\label{eq:LBphi}
\end{eqnarray}
where the covariant derivative is $D_\mu=\partial_\mu+i\Gamma_\mu$,
and $\delta_a$ introduces the mass splitting between the $\Sigma_c$
and the $\Sigma_c^*$. $\Gamma_\mu$ and $u_\mu$ denotes the chiral
connection (vector current) and the axial current of Goldstone boson
fields,
\begin{eqnarray}
  &&\Gamma_\mu=\frac{i}{2}[\xi^\dagger,\partial_\mu\xi]=-\frac{1}{4f_\pi^2}\epsilon^{abc}\tau^c(\phi^a\partial_\mu \phi^b)+\cdots,\nonumber\\
    &&u_\mu=\frac{i}{2}\{\xi^\dagger,\partial_\mu\xi\}=-\frac{1}{2f_\pi}\tau^a\partial_\mu \phi^a+\cdots\nonumber,\\
    &&\xi=\exp(i\phi/2f_\pi),\\
    &&\phi=\phi^a\tau^a=\sqrt{2}\begin{pmatrix}
        \frac{\pi^0}{\sqrt{2}}&\pi^+\\
        \pi^-&-\frac{\pi^0}{\sqrt{2}}
    \end{pmatrix},\nonumber
\end{eqnarray}
where $\phi$ denotes the Goldstone boson field, $\tau^a$ denotes the
Pauli matrices in $SU(2)$ flavor space, and $f_\pi=92$ MeV stands
for the pion decay constant.

Noticing that the total spin of light quarks in the superfield is
either 0 or 1, the spinor part of the superfield corresponds to the
spin of the heavy quark. Terms such as $\bar{\psi}^\mu
\myslash{u}\psi_\mu$ and $\bar{\psi}^\mu \sigma^{\nu\rho}u_\nu
v_\rho\psi_\mu$ are forbidden in the leading order, since the gamma
matrices break the HQSS. In principle, there are only two
independent coupling constants $g_1$ and $g_2$ in
Eq.~(\ref{eq:LBphi}). In previous works
\cite{Meng:2019ilv,Liu:2011xc}, they are usually related using quark
models, and $g_2$ can be evaluated from the
$\Sigma_c^{(*)}\rightarrow \Lambda_c\pi$ decay,
\begin{eqnarray}
  g_1=-1.47=-\sqrt{2}g_2,\qquad g_2=1.04.
\end{eqnarray}

The Lagrangians for the $\bar{D}^{(*)}$ sector can be constructed
similarly. The superfield $\tilde{H}$ for $\bar{D}^{(*)}$ reads
\begin{eqnarray}
  &&\tilde{H}=(\tilde{P}_\mu^*\gamma^\mu+i\tilde{P}\gamma_5)\frac{1-\myslash{v}}{2},\nonumber\\
  &&\bar{\tilde{H}}=\gamma^0H^\dagger\gamma^0=\frac{1-\myslash{v}}{2}(\tilde{P}_\mu^{*\dagger}\gamma^\mu+i\tilde{P}^\dagger\gamma_5),\\
  &&\tilde{P}=\begin{pmatrix}
      \bar{D}^0\\D^-
  \end{pmatrix}
  ,\qquad \tilde{P}_\mu^*=\begin{pmatrix}
      \bar{D}^{*0}\\D^{*-}
  \end{pmatrix}.\nonumber
\end{eqnarray}
Then the LO Lagrangian for the $\bar{D}^{(*)}\pi$ interaction reads
\begin{eqnarray}
  \mathcal{L}_{\tilde{H}\phi}&=-\langle(iv\cdot D \bar{\tilde{H}})\tilde{H}\rangle-\frac{1}{8}\delta\langle \bar{\tilde{H}}\sigma^{\mu\nu}\tilde{H}\sigma_{\mu\nu}\rangle+g\langle \bar{\tilde{H}}\myslash{u}\gamma_5\tilde{H}\rangle.
\end{eqnarray}
Apart from the OPE potential, contact terms are required to mimic
the short-range interactions. There are six independent terms in the
LO Lagrangian since the gamma matrices, which are related to the
spin of the heavy quark, are not allowed,
\begin{eqnarray}
  &\mathcal{L}_{\mathcal{B}H}&=C_1\,\langle\bar{\tilde{H}}\tilde{H}\rangle\Tr\left(\bar{\mathcal{B}}_1\mathcal{B}_1\right)+C_2\,\langle\bar{\tilde{H}}\gamma^\mu\gamma^5\tau^a\tilde{H}\rangle\Tr\left(\bar{\mathcal{B}}_1\tau^a\psi_\mu\right)+\text{h.c.}\\
  &&+C_3\,\langle\bar{\tilde{H}}\tilde{H}\rangle\Tr\left(\bar{\psi}^\mu\psi_\mu\right)+iC_4\,\epsilon_{\sigma\mu\nu\rho}v^\sigma\langle\bar{\tilde{H}}\gamma^\rho\gamma^5\tilde{H}\rangle\Tr\left(\bar{\psi}^\mu\psi^\nu\right)\\
  &&+C_5\,\langle\bar{\tilde{H}}\tau^a\tilde{H}\rangle\Tr\left(\bar{\psi}^\mu\tau^a\psi_\mu\right)+iC_6\,\epsilon_{\sigma\mu\nu\rho}v^\sigma\langle\bar{\tilde{H}}\gamma^\rho\gamma^5\tau^a\tilde{H}\rangle\Tr\left(\bar{\psi}^\mu\tau^a\psi^\nu\right),\label{eq:ct}
\end{eqnarray}
where $\langle\cdots\rangle$ stands for the trace in spinor space
and $\Tr(\cdots)$ stands for the trace in flavor space. The
$C_3(C_4)$ and $C_5(C_6)$ terms differ only by an isospin factor.
Since we focus on the $I=\frac{1}{2}$ case, there are only four
independent terms.

\section{Complex scaling method\label{sec:CSM}}

We use CSM to search for the possible bound states or resonances.
The CSM is an analytical extension of the Schr\"odinger equation,
proposed by Aguilar, Balslev and Combes
\cite{Aguilar:1971ve,Balslev:1971vb}. For a two-body scattering
process, it is equivalent to the non-relativistic LS equation.

We start from the Schr\"odinger equation in momentum space
\begin{eqnarray}
  E\phi_{l}(p)=\frac{p^2}{2m}\phi_{l}(p)+\int\frac{p'^2dp'}{(2\pi)^3}V_{l,l'}(p,p')\phi_{l'}(p')\label{eq:sch1},
\end{eqnarray}
where $l,l'$ are quantum numbers of the orbital angular momenta, and
$p$ denotes the momentum in the center-of-mass frame. The
corresponding LS equation reads
\begin{eqnarray}
  T(k',k;k_0)=V(k',k)+\int_0^\infty\frac{p^2dp}{(2\pi)^3}\frac{V(k',p)T(p,k;k_0)}{E_{k_0}-E_p+i0+},\label{eq:LS3}
\end{eqnarray}
where $E_{k_0}=\frac{k_0^2}{2m}$, $E_{p}=\frac{p^2}{2m}$ are the
non-relativistic kinetic energies.

With a complex scaling operation on Eq.~(\ref{eq:sch1}),
$p\rightarrow pe^{-i\theta},  \quad
\tilde{\phi}_l(p)=\phi_l(pe^{-i\theta})$, which will not change the
eigenenergy $E$, we derive the complex scaled Schr\"odinger equation
with a scaling angle $\theta$,
\begin{eqnarray}
  E\tilde{\phi}_{l}(p)&=\frac{p^2e^{-2i\theta}}{2m}\tilde{\phi}_{l}(p)+\int\frac{p'^2e^{-3i\theta}dp'}{(2\pi)^3}V_{l,l'}(pe^{-i\theta},p'e^{-i\theta})\tilde{\phi}_{l'}(p').\label{eq:sch2}
\end{eqnarray}

 An equivalent complex scaling operation in coordinate space will make the resonance wave functions convergent at $r\rightarrow\infty$. This can be roughly understood from the point of view of the asymptotic wave function
\begin{eqnarray}
  &\psi(r)&\xrightarrow{r\rightarrow\infty} f_l^+(k)e^{-ikr}+f_l^-(k)e^{ikr},\nonumber\\
  &&\xrightarrow{r\rightarrow r\exp(i\theta)}f_l^+(k)e^{-ikre^{i\theta}}+f_l^-(k)e^{ikre^{i\theta}},
\end{eqnarray}
where $f_l^\pm(k)$ are Jost functions. Resonances and bound states
correspond to the poles of the $T$ matrix, or the zeros of
$f_l^+(k)$. The first term vanishes and the second term converges
when $\text{Arg }k_{\text{res}}>-\theta$. However, we stress that
this is an inaccurate explanation since the asymptotic form is
obtained assuming $r$ is real, and in general, can not be extended
to the whole complex plane \footnote{For example, a transformation
$r\rightarrow-r$ will not make the bound states divergent or virtual
states convergent.}. A strict analytical form of the wave function
in coordinate space is obtained in Eq.~(\ref{eq:fourier}). As
explained in Sec.~\ref{sec:partialwidth}, whether the divergent term
shows up depends on the integral path and the poles of the wave
function.

\begin{figure}
  \includegraphics[width=75mm]{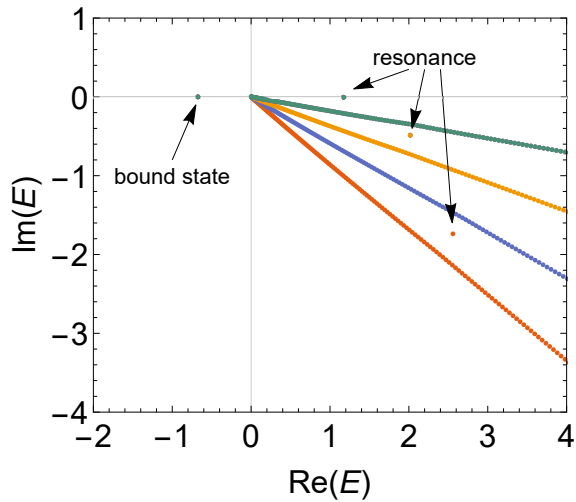}
  \caption{\label{fig:csm}A typical solution of the complex scaled Schr\"odinger equation. Eigenenergies are plotted on the complex plane. The continuum states line up due to the same arguments $\text{Arg} (E)=-2\theta$. Points with different colors stand for eigenenergies solved under different $\theta$. With a Hermitian Hamiltonian, the bound states lie on the negative real axis, while the resonances lie on the fourth quadrant, and appear only when $|\text{Arg} (E)|<2\theta$. }
\end{figure}

A typical distribution of the eigenenergies solved by the CSM is
shown in Fig.~\ref{fig:csm}. The continuum states line up and rotate
as the complex scaling angle $\theta$ varies. Poles are isolated
from the continuum states. The region between the continuum line and
$+x$-axis corresponds the second Riemann sheet and is where the
resonances lie. The rest of the complex plane corresponds to the 1st
Riemann sheet and is where the bound states lie.

In the coupled-channel cases, only the bound states below the lowest
channel can be directly solved in the normal Schr\"odinger equation
(Eq.~(\ref{eq:sch1})). For a "bound state" coupling to an open
channel with a lower threshold, its wave function of the lower
channel will be divergent, and thus can not be solved. Poles of this
type lie on the 1st Riemann sheet of the higher channel, and the
second Riemann sheet of the lower channel. They are called the
quasi-bound states, Feshbach-type resonances or unstable bound
states. Using CSM, both their energies and widths can be solved
directly from Eq.~(\ref{eq:sch2}). For a more precise classification
of the poles, see Ref.~\cite{Badalian:1981xj}.

\section{Complex scaled wave function and partial
width \label{sec:partialwidth}}

Since the complex scaled Schr\"odinger equation is equivalent to the
LS equation, we can dig out the information of the $T$ matrix at the
resonance energy $E_{\text{res}}$ from the corresponding resonance
wave function. We will present an approach to calculate the residues
of the $T$ matrix, which corresponds to the partial widths of the
states under the narrow resonance approximation. There are several
previous works dealing with the partial widths using CSM
\cite{Noro-Taylor,Masui:1999vjf}. Our approach is derived in
momentum space without extra approximations.

The complex scaled wave function $\tilde{\psi}(r)$ solved from
Eq.~(\ref{eq:sch2}) is indeed an analytical extension of the real
wave function $\psi(r)$ solved from Eq.~(\ref{eq:sch1}). They are
associated through the Schr\"odinger equation,
\begin{eqnarray}
  \langle k|\hat{T}+\hat{V}|\phi\rangle=E_R\langle k|\phi\rangle,
\end{eqnarray}
where $\hat{T}$, $\hat{V}$ denotes the kinetic energy and the
potential, respectively, and the $E_R=M-i\frac{\Gamma}{2}$ is the
resonance energy.

In the momentum presentation, we obtain
\begin{eqnarray}
  \frac{\bm{k}^2}{2m}\phi(\bm{k})+\int\frac{d^3\bm{p}}{(2\pi)^3}V(\bm{k},\bm{p})\phi(\bm{p})=E_R\phi(\bm{k}),
\end{eqnarray}
or
\begin{eqnarray}
  \phi(\bm{k})=\frac{1}{E_R-\frac{\bm{k}^2}{2m}}\int\frac{d^3\bm{p}}{(2\pi)^3}V(\bm{k},\bm{p})\phi(\bm{p}),
\end{eqnarray}
where $\bm{k}$ can be set to be any complex value, while $\bm{p}$ is
always real as long as we carry out the integral along the real
axis. Then we can extend the wave function to the complex plane once
we obtain the wave funtion on the position real axis.

Furthermore, we can employ a complex scaling operation and derive
\begin{eqnarray}
  \phi(\bm{k})=\frac{1}{E_R-\frac{\bm{k}^2}{2m}}\int\frac{d^3\bm{p}}{(2\pi)^3}e^{-3i\theta}V(\bm{k},\bm{p}e^{-i\theta})\tilde{\phi}(\bm{p}),\label{eq:extend}
\end{eqnarray}
where $\tilde{\phi}(\bm{p})=\phi(\bm{p}e^{-i\theta})$.

The complex scaling operation is not always feasible since it
requires
$\lim_{p\rightarrow\infty}p^3V_{l,l'}(k,p)\phi(p)\rightarrow 0$.
Here we assume it to be true since the potential usually contains a
cutoff at large momenta. But it constrains the range of the complex
scaling angle $\theta$, because the potential $V_{l,l'}(k,p)$ is
always accompanied by some non-analytical behaviors (unless it is a
constant). For example, a monopole or dipole regulator
$1/(q^2+\Lambda^2)^n$ introduces a left-hand cut to the potential
after the partial wave expansion, and an exponential regulator
$\exp(-q^n/\Lambda^n)$ introduces a singularity at infinity, which
sets the constraint $\theta<\pi/(2n)$ to avoid divergence.

From Eq.~(\ref{eq:extend}), we see that $\phi(k)$ diverges at
$k=\pm\sqrt{2mE_R}$. This applies to both the bound states and
resonances in the scattering problems, in which the potential
satisfies
\begin{eqnarray}
  \lim_{r\rightarrow\infty} r^2V(r)\rightarrow 0.
\end{eqnarray}

The poles of the wave function result in discontinuity of the
integral. (See Fig.~\ref{fig:int} in Sec.~\ref{chap:3bodies} as an
example.) In Eq.~(\ref{eq:sch2}) and Eq.~(\ref{eq:extend}), the
integral paths above and below the pole differ by the residue of the
pole. Whether we take into account the contribution of the pole in
the integral determines the type of the state. For a resonance on
the second Riemann sheet, the integral should be performed below the
pole.

Then we focus on the wave function in coordinate space, which is
related to the wave function in momentum space through the Fourier
transformation
\begin{eqnarray}
  \psi_l(r)=\int_0^\infty \frac{4\pi p^2dp}{(2\pi)^3}e^{-3i\theta}\phi_l(pe^{-i\theta})i^l j_l(pre^{-i\theta}),
\end{eqnarray}
where $j_l$ is the $l$-th spherical Bessel function, which is
divergent at $p\rightarrow\infty$ when $0<\theta<\pi$. If we change
the integral path to $\theta=0$ then we have to add the residue of
the pole to compensate for the discontinuity
\begin{eqnarray}
  &\psi_l(r)&=\int_0^\infty \frac{4\pi p^2}{(2\pi)^3}\phi_l(p)i^l j_l(pr)dp+2\pi i \text{Res}\{\frac{4\pi p^2}{(2\pi)^3}\phi_l(p)i^l j_l(pr)\}\Big|_{p=k_R}\nonumber\\
  &&=\int_0^\infty \frac{4\pi p^2}{(2\pi)^3}\phi_l(p)i^l j_l(pr)dp+i^{l+1}\frac{k_R^2}{\pi}j_l(k_Rr)\lim_{p\rightarrow k_R}(p-k_R)\phi_l(p),\label{eq:fourier}
\end{eqnarray}
where $k_R=\sqrt{2mE_R}$.

The first term in Eq.~(\ref{eq:fourier}) is convergent at
$r\rightarrow\infty$, and the second term gives the asymptotic
behavior of $\psi_l(r)$ at $r\rightarrow\infty$,
\begin{eqnarray}
  \psi_l(r)\rightarrow \frac{i^lk_R}{2\pi}\lim_{p\rightarrow k_R}(p-k_R)\phi_l(p)\frac{e^{i(k_Rr-\pi l/2)}}{r},
\end{eqnarray}
which diverges when $k_R$ is not real. If we employ a transformation
$r\rightarrow re^{i\theta}$, then we have to replace $p\rightarrow
pe^{-i\theta}$ in the integral to keep $j_l(pr)$ convergent at
$r\rightarrow\infty$. When the integral path passes the pole, the
second term vanishes and the wave function becomes convergent.

In multichannel cases, the wave function of the $j$-th channel
satisfies
\begin{eqnarray}
  \psi_{l,j}(r)\rightarrow \frac{i^lk_{R,j}}{2\pi}\lim_{p\rightarrow k_{R,j}}(p-k_{R,j})\phi_{l,j}(p)\frac{e^{i(k_{R,j}r-\pi l/2)}}{r}.
\end{eqnarray}
where $k_{R,j}=\sqrt{2m_j(E_R-E_{\text{th},j})}$, $m_j$ and
$E_{\text{th},j}$ are the (reduced) mass and the threshold of the
$i$-th channel.

\begin{figure}
  \centering
  \includegraphics[width=80mm]{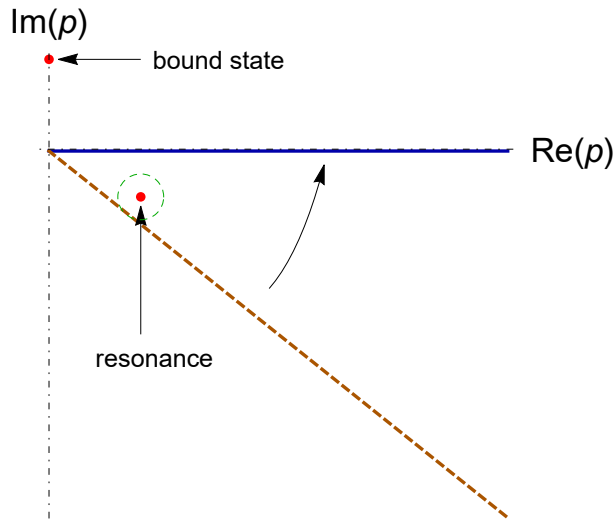}
\caption{The differences between the poles on the 1st and 2nd
Riemann sheets. Integrals along the brown dashed line and the blue
solid line are equal for the bound states, but different for the
resonances. The correct choice of the integral paths for the
resonances is the brown dashed line, which corresponds to the
equation on the 2nd Riemann sheet. For the resonances, if we change
the integral path to the blue solid one, then we need an additional
integral path along the green dashed circle for compensation, which
corresponds to the residue of the pole.\label{fig:int_res}}
\end{figure}

We stress that Eq.~(\ref{eq:fourier}) applies only to the poles on
the 2nd Riemann sheet. For the poles on the 1st Riemann sheet, the
pole will not cross the integral path when $\theta$ varies to zero.
As shown in Fig.~\ref{fig:int_res}, for a pole on the 1st Riemann
sheet, the integral path can be rotated continuously from $\theta$
to 0. So the second term in Eq.~(\ref{eq:fourier}) vanishes. The
wave function tends to 0 as $r\rightarrow\infty$. In other words,
the state will not decay into this channel. When calculating the
partial widths, we only need to consider the channels in which the
pole is on the 2nd Riemann sheet, no matter whether the resonance
energy is above or below the threshold. Notably, $k_{R,j}$ is
usually complex and $k_{R,j}$ in different channels can be quite
different.

The coefficients of the spherical wave in different channels
correspond to the component proportions of the out-going state. It
represents the amplitudes of decaying to different final states and
is related to the branching ratios of the resonance. It is not the
modulus of the wave function but the residue of the wave function at
the resonance energy that determines the branching ratios. Following
the formula derived by Moiseyev and Peskin \cite{Moiseyev-Peskin},
we obtain
\begin{eqnarray}
  \frac{\Gamma_{1}}{\Gamma_{2}}=\left|\frac{k_{R,1}/\mu_1}{k_{R,2}/\mu_2}\right| \left|\frac{k_{R,1}\lim_{p\rightarrow k_{R,1}}(p-k_{R,1})\phi_{1}(p)}{k_{R,2}\lim_{p\rightarrow k_{R,2}}(p-k_{R,2})\phi_{2}(p)}\right|^2.
\end{eqnarray}

Directly calculating the residue of the wave function may yield a
large numerical error, so we use an alternative approach. Noticing
that the residue of $\phi(p)$ at $p=k_R$ can be calculated through
Eq.~(\ref{eq:extend}) by letting $k\rightarrow k_R$, we finally
derive
\begin{eqnarray}
  \frac{\Gamma_{1}}{\Gamma_{2}}=\left|\frac{k_{R,1}\mu_1}{k_{R,2}\mu_2}\right| \left|\frac{\langle k_{R,1}|\hat{V}|\phi\rangle}{\langle k_{R,2}|\hat{V}|\phi\rangle}\right|^2,\label{eq:branchingratio}
\end{eqnarray}
where
\begin{eqnarray}
  \langle k_{R,j}|\hat{V}|\phi\rangle=\int\frac{p^2dp}{(2\pi)^3}e^{-3i\theta}V_{jm}(k_{R,j},pe^{-i\theta})\tilde{\phi}_m(p),
\end{eqnarray}
where $j$, $m$ are the channel labels.

If we choose the normalization condition defined by c-product
\cite{Romo:1968tcz}, which reads
\begin{eqnarray}
  (\phi|\phi)=\int \frac{d^3\bm{k}}{(2\pi)^3}\phi(k)^2=\int \frac{d^3\bm{k}}{(2\pi)^3}e^{-3i\theta}\phi(ke^{-i\theta})^2=1,
\end{eqnarray}
then the expression above is exactly the residue of the $S$ matrix
or $T$ matrix,
\begin{eqnarray}
  \text{Res } |S_{jj}(E)|\Big|_{E=E_R}=\left|\frac{\mu_jk_{R,j}}{4\pi^2} \langle k_{R,j}|\hat{V}|\phi\rangle^2\right|,
\end{eqnarray}
where $j$ stands for the $j$-th channel.

Compared with the LS equation, CSM gives the same information of the
positions and residues of the poles, but in a more direct way, since
one does not have to search for the poles.

For the narrow resonances away from thresholds, the residue of the
$S$ matrix corresponds to the partial width of the resonance,
\begin{eqnarray}
  S_{ij}(E)\approx \delta_{i,j}-\frac{i\sqrt{\Gamma_i\Gamma_j}}{E-(M-\frac{i}{2}\Gamma)}.\label{eq:Smatrix}
\end{eqnarray}

For the resonances near the threshold, Eq.~(\ref{eq:Smatrix}) does
not hold anymore. The partial widths can be still well defined as
the residues of the $S$ matrix, but they no longer add up to the
total width. In this work, we only use Eq.~(\ref{eq:branchingratio})
to calculate the branching ratio, which is independent of the choice
of the normalization conditions.

\section{Effects of the three-body threshold\label{chap:3bodies}}

In line with Ref.~\cite{Lin:2022wmj}, the on-shell intermediate
$DD(\bar{D})\pi$ state plays an important role in the width of the
$T_{cc}^+$ and $\chi_{c1}(3872)$. Since the mass splitting between
$\Lambda_c$ and $\Sigma_c^{(*)}$ is larger than the pion mass, we
include the effect of the three-body intermediate states in the OPE
potential in this work.

Still, we retain the 0-th component of the transferred momentum
$q=p-p'$ for the OPE potential
\begin{eqnarray}
  V_{1\pi}=-C\frac{(S_1\cdot q)(S_2\cdot q)}{q_0^2-\bm{q}^2-m_\pi^2+i0+},\label{eq:OPE1}\\.
\end{eqnarray}
where $S_1$ and $S_2$ are the spin operators of
$\Sigma_c^{(*)}(\Lambda_c)$ and $\bar{D}^{(*)}$, and $p,p'$
represent the center-of-mass momentum of the initial and final
state, respectively. $C$ is a constant depending on the isospin and
coupling constants.

\begin{figure}
  \centering
  \begin{minipage}{0.45\linewidth}
    \includegraphics[height=30mm]{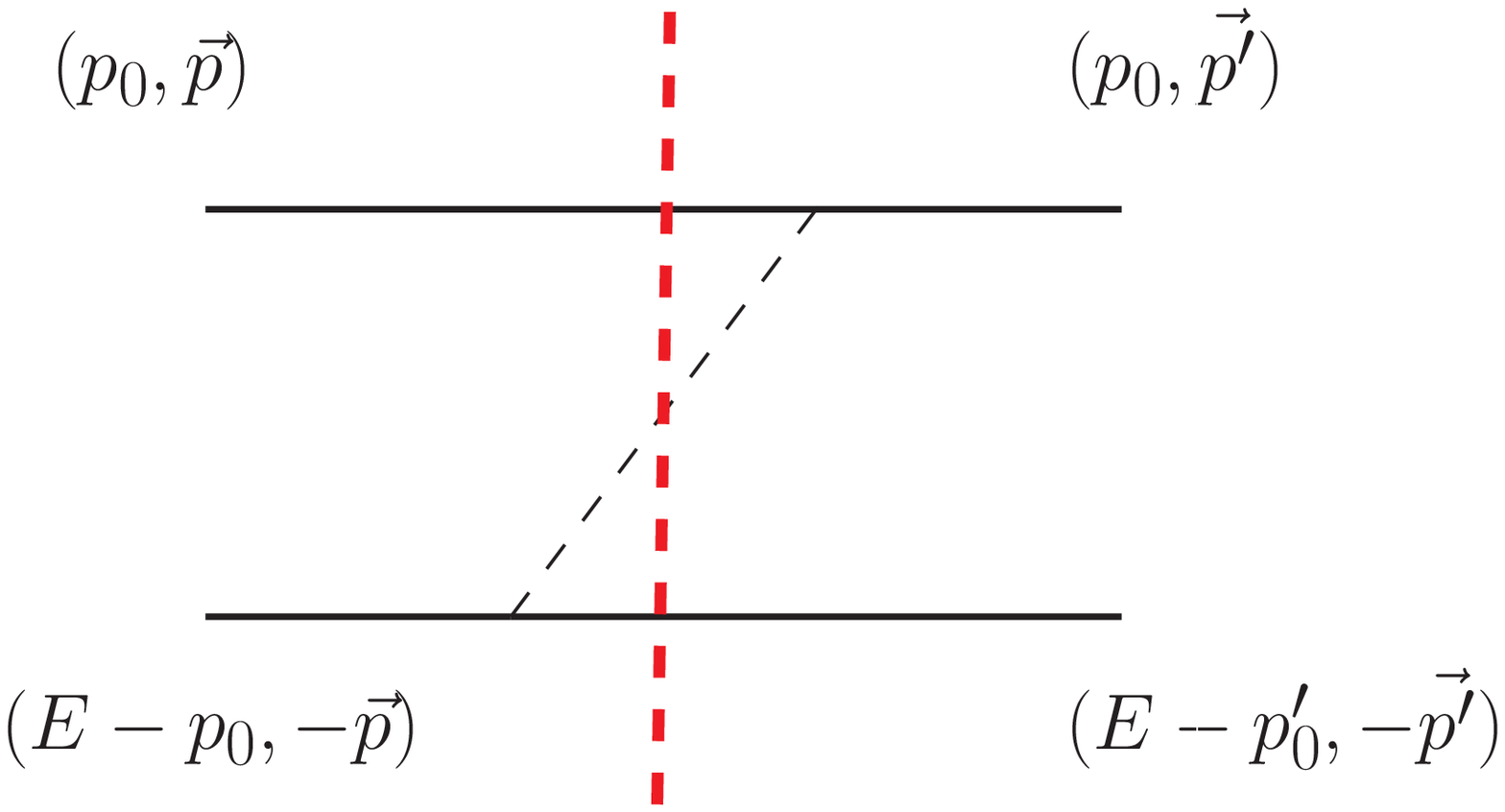}
  \end{minipage}
  \begin{minipage}{0.45\linewidth}
    \includegraphics[height=30mm]{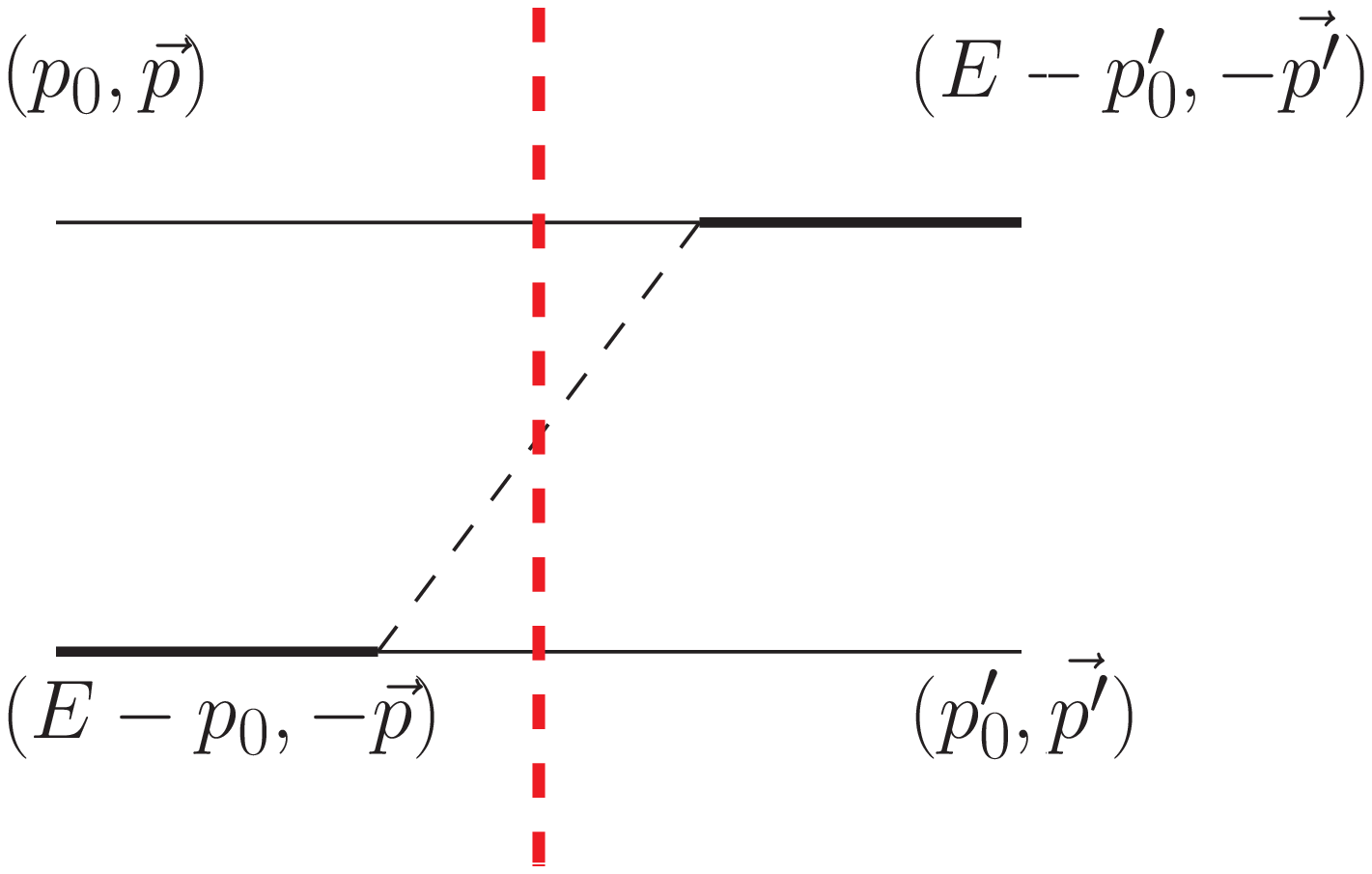}
  \end{minipage}
  \caption{\label{fig:cutOPE}The one-pion-exchange diagram. Left: direct diagram ($P_c$); Right: cross diagram ($T_{cc}^+$). $E$ denotes the center-of-mass energy. The on-shell intermediate state contributes to the imaginary part.}
\end{figure}

Different from the $DD^*$ system, the OPE potential in
$\Sigma_c^{(*)}(\Lambda_c)\bar{D}^{(*)}$ arises from a direct
diagram rather than a cross diagram (see Fig.~\ref{fig:cutOPE}). We
choose the $q^0$ as the form
\begin{eqnarray}
  q^0=\sqrt{\vec{p'}^2+m_3^2}-\sqrt{\vec{p}^2+m_1^2}.
\end{eqnarray}
We start from the 4-dimensional equation
\begin{eqnarray}
  T_{ij}(\bm{p}',\bm{p};p'_0,p_0,E)=V_{ij}(\bm{p}',\bm{p};p'_0,p_0)+\int\frac{d^4 l}{(2\pi)^4}V_{ik}(\bm{p}',\bm{l};p'_0,l_0)\mathcal{G}_k(l;E)T_{kj}(\bm{l},\bm{p};l_0,p_0,E).\label{eq:LS4}
\end{eqnarray}
To avoid ambiguity, we write the 3-momentum $\bm{p}',\bm{p}$ and
energy $p'_0,p_0$ separately. $i,j,k$ are the channel labels. $E$ is
the total energy and is conserved in the initial, intermediate and
final states. In general, the $T$ matrices need not be on-shell,
namely $p'_0,p_0,E$ can be set to any value regardless of
$\bm{p}',\bm{p}$. But for the on-shell $T$ matrices, all momenta are
on-shell and the energies sum up to $E$, either in the initial or
final states.

Then we perform the integral on $l^0$ using the residue theorem. The
propagator of the $i$-th channel reads
\begin{eqnarray}
  \mathcal{G}_k(l;E)=\frac{i}{(l^2-M_{k1}^2+i\epsilon)[(P-l)^2-M_{k2}^2+i\epsilon]},
\end{eqnarray}
where $M_{k1}$, $M_{k2}$ denote the masses of the particles in the
$k$-th channel. $P=(E,0,0,0)$ is the total 4-momentum.

Then the poles read
\begin{eqnarray}
  \varepsilon^{\pm}_1=\mp\left(\sqrt{\bm{l}^2+M_{k1}^2}-i\epsilon\right)\approx \mp M_{k1},\qquad \varepsilon_2^{\pm}=E\mp\left(\sqrt{\bm{l}^2+M_{k2}^2}-i\epsilon\right)\approx E\mp M_{k2}.
\end{eqnarray}

\begin{figure}
  \centering
  \includegraphics{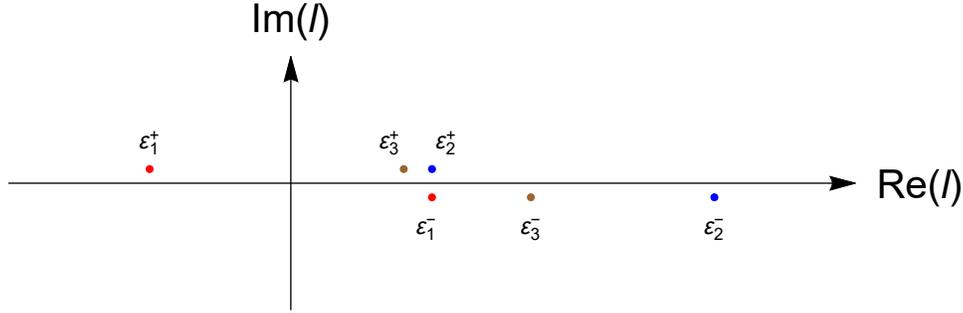}
\caption{\label{fig:poleposition}The poles arise from the integral
in Eq.~(\ref{eq:LS4}). The positions of $\varepsilon_3^{\pm}$ depend
on the mass difference of the initial and final particles. In this
figure, the major contributions arise from $\varepsilon_1^{-}$,
$\varepsilon_2^+$ and $\varepsilon_3^+$. We can perform the integral
in the lower half plane and consider only the residue of
$\varepsilon_1^{-}$.}
\end{figure}

The relative position of the poles are shown in
Fig.~\ref{fig:poleposition}. There is also a pole in
$V_{ik}(\bm{p}',\bm{l};p'_0,l_0)$, which can be derived from
Eq.~(\ref{eq:OPE1})
\begin{eqnarray}
  \varepsilon^{\pm}_3=p'_0\mp\left(\sqrt{(\bm{l}-\bm{p}')^2+m_{\pi}^2}-i\epsilon\right)\approx M_{i1}\mp m_\pi
\end{eqnarray}

The possible poles of the $T$ matrices are not considered. For
illustration, we regard Eq.~(\ref{eq:LS4}) as an analytical
extension of $T$ on $p'_0$-plane (like Eq.~(\ref{eq:extend})) by
setting the $\bm{p}',p_0$ to complex values while the integral on
$l$ is performed along the real axis. Obviously, there are no poles
of $T$ on the $p'_0$-plane other than those of $V$. In fact, we can
solve the $T$ matrix formally $T=[1-VG]^{-1}V$. If we fix the $p$
and $E$ to certain values and discretize the $p'$ or $l$, then the
$T$ matrix will reduce to a "vector". The poles correspond to the
zeros of the determinant $|1-VG|$, where every term of the
discretized $T$ "vector" is divergent. I.e., even a half-on-shell
$T(p',p;E)$ matrix ($p$ is on-shell while $p'$ is not) is divergent.
The pole structure does not appear on the $p'_0$-plane but is
related to the variable $E$.

Since we are discussing low energy physics, we suppose the integral
on $\bm{l}$ is regulated in a certain way, so the integral is
convergent and 3-momentum is small compared to the masses of the
charmed hadrons. Then we can estimate the contributions of the
poles.

In general, the far-away poles contribute little to the integral,
and we only count the poles with the other poles in their
neighborhood. For instance, $\varepsilon_1^-$ and $\varepsilon_2^+$
are close to each other since $E\approx M_{k1}+M_{k2}$ in the
non-relativistic limit, from which the non-relativistic form is
deduced,
\begin{eqnarray}
  &\int\frac{d^4 l}{(2\pi)^4}V_{ik}(\bm{p}',\bm{l};p'_0,l_0)\mathcal{G}_k(l;E)T_{kj}(\bm{l},\bm{p};l_0,p_0,E)&\rightarrow -2\pi \int\frac{d^3 l}{(2\pi)^4}\frac{V_{ik}(\bm{p}',\bm{l};p'_0,\varepsilon_2^+)T_{kj}(\bm{l},\bm{p};\varepsilon_2^+,p_0,E)}{(\varepsilon_2^+-\varepsilon_2^-)(\varepsilon_2^+-\varepsilon_1^-)(\varepsilon_2^+-\varepsilon_1^+)},\nonumber\\
  &&= \int\frac{d^3 l}{(2\pi)^3}\frac{V_{ik}(\bm{p}',\bm{l};p'_0,\varepsilon_2^+)T_{kj}(\bm{l},\bm{p};\varepsilon_2^+,p_0,E)}{4E\sqrt{M_{k2}^2+\bm{l}^2}(\sqrt{M_{k2}^2+\bm{k}_0^2}-\sqrt{M_{k2}^2+\bm{l}^2})},\label{eq:LSr}
\end{eqnarray}
where the center-of-mass momentum $\bm{k}_0$ satisfies
$E=\sqrt{M_{k1}^2+\bm{k}_0^2}+\sqrt{M_{k2}^2+\bm{k}_0^2}$. After
applying the non-relativistic reduction and considering the
normalization constants, we obtain Eq.~(\ref{eq:LS3}).

The poles $\varepsilon_1^{+}$ and $\varepsilon_2^-$ can be
dropped\footnote{For energy-independent potentials, we usually sum
up the residues of two poles $\varepsilon_1^+$ and
$\varepsilon_2^+$, but in fact the former is small.} because there
is a $(E+M_{k1}+M_{k2})$ in the denominator, and
$V_{ik}(\bm{p}',\bm{l};p'_0,p_0)$ is small when $p_0$ deviates from
$p'_0$. However, the poles $\varepsilon_3^{\pm}$ have a considerable
contribution when $\varepsilon_3^{\pm}$ gets close to
$\varepsilon_2^+$ by accident. The condition reads
\begin{eqnarray}
  |M_{i1}-M_{k1}|\sim m_\pi,
\end{eqnarray}
which turns out to be true for the $\Sigma_c-\Lambda_c$ or $D^*-D$
systems.

Instead of directly calculating the residue of
$\varepsilon_3^{\pm}$, we select an appropriate integral contour to
include only one pole. For example, if $M_{k1}-M_{i1}\sim m_\pi$, we
perform the integral in the upper half plane and consider only
$\varepsilon_2^+$. Subsequently, we obtain the 3-dimensional LS
equation
\begin{eqnarray}
  T_{ij}(\bm{p}',\bm{p};p'_0,p_0,E)=V_{ij}(\bm{p}',\bm{p};p'_0,p_0)+\int\frac{d^3 l}{(2\pi)^3}V_{ik}(\bm{p}',\bm{l};p'_0,\varepsilon_2^+(\bm{l}))G_k(\bm{l};E)T_{kj}(\bm{l},\bm{p};\varepsilon_2^+(\bm{l}),p_0,E),\label{eq:LS4}
\end{eqnarray}
where $G_k(\bm{l};E)$ is a 3-dimensional propagator, as seen in
Eq.~(\ref{eq:LS3}). In different channels, the choice of
$\varepsilon_2^+$ and $\varepsilon_1^-$ can be determined
independently. Notably, the $T$ matrix in the left and right sides
must have the same form to ensure the LS equation to be an iterative
equation. Then $p'_0$ must be set to $\varepsilon_2^+(\bm{p}')$
\footnote{One may notice that $V_{ij}(\bm{p}',\bm{p};p'_0,p_0)$ is
different from
$V_{ik}(\bm{p}',\bm{l};p'_0,\varepsilon_2^+(\bm{l}))$. This will not
cause problems since the poles depend only on the $V$ in the
integral.}.

Then an appropriate choice of $q_0$ in the OPE potential reads
\begin{eqnarray}
  q_0=\left\{\begin{array}{cc}
    \varepsilon_2^+(\bm{p}')-\varepsilon_2^+(\bm{p})=\sqrt{\bm{p}^2+M_{k2}^2}-\sqrt{\bm{p}'^2+M_{i2}^2},& M_{k1}-M_{i1}\sim m_\pi,\\
    \varepsilon_1^-(\bm{p}')-\varepsilon_1^-(\bm{p})=\sqrt{\bm{p}'^2+M_{k1}^2}-\sqrt{\bm{p}^2+M_{i1}^2},& M_{i1}-M_{k1}\sim m_\pi.
  \end{array}\right.
\end{eqnarray}
One can use either of them if the mass difference is far away from
the pion mass. If repeating the analysis of the cross diagram in the
$DD^*$ system, one will find the appropriate choice of $q_0$ is
always $\pm(E-\sqrt{\bm{p}^2+M_D^2}-\sqrt{\bm{p}'^2+M_D^2})$, as
presented in Ref.~\cite{Lin:2022wmj}, which indicates that it is the
$DD\pi$ intermediate state, rather than $D^*D^*\pi$, that is
important.

\begin{figure}
  \includegraphics[width=75mm]{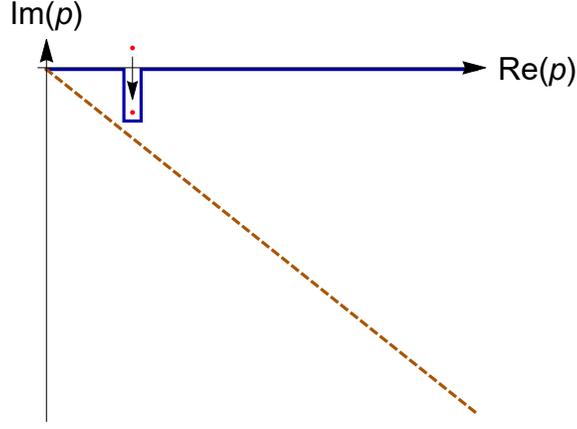}
\caption{\label{fig:int}The integral path from $0$ to $\infty$ in
the complex $p$-plane. The red point denotes the pole of the OPE
potential located at $p=\sqrt{p_0^2-m_\pi^2}$. When the pole passes
across the positive real axis, we need to change the integral path
to maintain the analytical continuity (blue solid line). Instead, we
can carry out a complex scaled integral (brown dashed line) to deal
with the pole.}
\end{figure}

When $q_0>m_\pi$, there is a singularity in the OPE potential, which
results in a right-hand (unitary) cut. It is related to the on-shell
three-body intermediate state according to the optical theorem. As
shown in Fig.~\ref{fig:int}, we use a nonzero complex scaling angle
$\theta$ to skip the singularity and perform the integral in the 2nd
Riemann sheet with respect to the three-body threshold.

\section{Effective potential\label{sec:Veff}}

Compared with the Born approximation in the scattering, the
effective potential is related to the Feynman amplitude of the
two-particle-irreducible diagrams
\begin{eqnarray}
  V=-\frac{1}{4}\mathcal{M},
\end{eqnarray}
where the factor $-\frac{1}{4}$ differs from the usual
$-\prod_i\frac{1}{\sqrt{2M_i}}$ because of the normalization of the
heavy meson and baryon fields.

We adopt a Gaussian regulator to suppress the potential $V$ at large
momentum $p$, $p'$, which reads
\begin{eqnarray}
  \mathcal{F}(\bm{p},\bm{p}')=\exp\left[-(\bm{p}^2+\bm{p}'^2)/\Lambda^2)\right].\label{eq:regulator}
\end{eqnarray}
We demand $\mathcal{F}(\bm{p},\bm{p'})\rightarrow 0$ when
$\bm{p},\bm{p}'\rightarrow\infty$ before and after the rotation in
the complex plane to ensure that the Schr\"odinger equation can be
solved numerically, which constrains the rotating angle
$\theta<\pi/4$.

A coupled-channel calculation is performed to investigate the mass
and width of the $P_c$ states. The channels considered are listed in
Table~\ref{tab:channels}. In this work, only the $I=\frac{1}{2}$
channels are considered. The masses and widths of the particles are
listed in Table~\ref{tab:mass}. We use an average mass for an
isospin multiplet. We do not consider the effect of their widths,
although they contribute to the width of the $P_c$ states. Apart
from the $\Sigma_c^*$, the width of the other charmed hadrons is no
more than 2 MeV. Since the $P_c$ states are below the nearest
thresholds, the off-shell width of $\Sigma_c$ is smaller than its
on-shell value. In this work, we leave it as systematic
uncertainties.

\begin{table}
  \renewcommand\arraystretch{1.4}
  \caption{\label{tab:channels}The channels considered in the $\Sigma_c^{(*)}(\Lambda_c)\bar{D}^{(*)}$ systems ($I=\frac{1}{2}$).}
  \begin{ruledtabular}
  \begin{tabular}{cccccc}
    &1&2&3&4&5\\
    \hline
    $J^P=\frac{1}{2}^-$ & $\Lambda_c\bar{D}$ & $\Lambda_c\bar{D}^*$ & $\Sigma_c\bar{D}$ & $\Sigma_c\bar{D}^*$ & $\Sigma_c^*\bar{D}^*$\\
    $J^P=\frac{3}{2}^-$&$\Lambda_c\bar{D}^*$&$\Sigma_c\bar{D}^*$&$\Sigma_c^*\bar{D}$&$\Sigma_c^*\bar{D}^*$\\
  \end{tabular}
  \end{ruledtabular}
  \end{table}

  \begin{table}
    \renewcommand\arraystretch{1.3}
    \caption{\label{tab:mass} The masses and widths of the charmed baryons and mesons (MeV). "--" means long-life particles whose width can be ignored. \cite{ParticleDataGroup:2022pth}}
    \begin{ruledtabular}
    \begin{tabular}{cccccc}
      baryons&mass&width&mesons&mass&width\\
      \hline
      $\Lambda_c^+$ & 2286.46 & -- & $D^-$ & 1869.66 & --\\
      $\Sigma_c^{++}$&2453.97&1.89&$\bar{D}^0$&1864.84&--\\
      $\Sigma_c^{+}$ & 2452.65 & 2.3 & $D^{*-}$ & 2010.26 & 8.34$\times10^{-2}$\\
      $\Sigma_c^{0}$ & 2453.75 & 1.83 &$\bar{D}^{*0}$&2006.85& $<2.1$\\
      $\Sigma_c^{*++}$ &2518.41 & 14.78 &&&\\
      $\Sigma_c^{*+}$ & 2517.4 & 17.2 &&&\\
      $\Sigma_c^{*0}$ & 2518.48 & 15.3 &&& \\
    \end{tabular}
    \end{ruledtabular}
    \end{table}

The LO contact terms have been investigated in many previous works
\cite{Du:2019pij,Meng:2019ilv,Xiao:2019aya,Burns:2022uiv}. We adopt
the notations in Ref.~\cite{Burns:2022uiv} and rewrite the contact
terms in Table~\ref{tab:j1/2} and \ref{tab:j3/2} for simplification.
The extra $i$ arises from the relative phase of $\bar{D}$ and
$\bar{D}^*$ in the definition of the superfield. The matrix is
Hermitian and the lower half of the table is omitted. The constant
$A$ and $B$ correspond to $C_1$ and $C_2$ in Eq.~(\ref{eq:ct}).
$C_a$ corresponds to $C_3$ and $C_5$. $C_b$ corresponds to $C_4$ and
$C_6$. We determine the LECs by fitting the masses and widths of the
$P_c$ states.

\begin{table}
  \renewcommand\arraystretch{1.4}
  \caption{\label{tab:j1/2}The contact terms for the $J^P=\frac{1}{2}^-$ channels.}
  \begin{ruledtabular}
  \begin{tabular}{cccccc}
      $1 / 2^{-}$ & $\Lambda_{c} D$ & $\Lambda_{c} \bar{D}^{*}$ & $\Sigma_{c} \bar{D}$ & $\Sigma_{c} \bar{D}^{*}$ & $\Sigma_{c}^{*} \bar{D}^{*}$ \\
      \hline
      $\Lambda_{c} \bar{D}$ & $A$ & 0 & 0 & $\sqrt{3} iB$ & $\sqrt{6} iB$ \\
      $\Lambda_{c} \bar{D}^{*}$ & & $A$ & $-\sqrt{3} iB$ & $-2 B$ & $\sqrt{2} B$ \\
      $\Sigma_{c} \bar{D}$ & & & $C_{a}$ & $\frac{2}{\sqrt{3}} iC_{b}$ & $-\sqrt{\frac{2}{3}} iC_{b}$ \\
      $\Sigma_{c} \bar{D}^{*}$ & & & & $C_{a}-\frac{4}{3} C_{b}$ & $-\frac{\sqrt{2}}{3} C_{b}$ \\
      $\Sigma_{c}^{*} \bar{D}^{*}$ & & & & & $C_{a}-\frac{5}{3} C_{b}$ \\
    \end{tabular}
    \end{ruledtabular}
\end{table}

\begin{table}
  \renewcommand\arraystretch{1.4}
  \caption{\label{tab:j3/2}The contact terms for the $J^P=\frac{3}{2}^-$ channels.}
  \begin{ruledtabular}
  \begin{tabular}{ccccc}
    $3 / 2^{-}$ & $\Lambda_{c} \bar{D}^{*}$ & $\Sigma_{c}^{*} \bar{D}$ & $\Sigma_{c} \bar{D}^{*}$ & $\Sigma_{c}^{*} \bar{D}^{*}$ \\
    \hline
    $\Lambda_{c} \bar{D}^{*}$ & $A$ & $\sqrt{3} iB$ & $B$ & $\sqrt{5} B$ \\
    $\Sigma_{c}^{*} \bar{D}$ & & $C_{a}$ & $\frac{iC_{b}}{\sqrt{3}}$ & $\sqrt{\frac{5}{3}} iC_{b}$ \\
    $\Sigma_{c} \bar{D}^{*}$ & & & $C_{a}+\frac{2}{3} C_{b}$ & $-\frac{\sqrt{5}}{3} C_{b}$ \\
    $\Sigma_{c}^{*} \bar{D}^{*}$ & & &  & $C_{a}-\frac{2}{3} C_{b}$\\
    \end{tabular}
    \end{ruledtabular}
\end{table}

Since we consider only the S-wave interactions, the OPE potential
can be written in an equivalent form,
\begin{eqnarray}
  V_{1\pi}=-C\frac{(S_1\cdot q)(S_2\cdot q)}{q_0^2-\bm{q}^2-m_\pi^2+i0+}\rightarrow -C\frac{\frac{1}{3}\bm{q}^2(S_1\cdot S_2)}{q_0^2-\bm{q}^2-m_\pi^2+i0+},
\end{eqnarray}
where $S_1$ and $S_2$ are the vectors related to the spin operators
of the $\Sigma_c^{(*)}(\Lambda_c)$ or the $\bar{D}^{(*)}$.

In Table \ref{tab:ope1/2} and Table \ref{tab:ope3/2}, we list the
coefficients $C_{ij}$ of the OPE potentials. $q_0$ for the different
channels is determined to be the energy difference of
$\Sigma_c^{(*)}$ or $\Lambda_c$ in the initial and final states. The
coefficient matrix is symmetric and the lower half of the table is
omitted. $C_{ij}$ is defined as
\begin{eqnarray}
  V_{1\pi,ij}=C_{ij}\frac{1}{f_\pi^2}\frac{\bm{q}^2}{-q_0^2+\bm{q}^2+m_\pi^2-i0+}.
\end{eqnarray}

\begin{table}
  \renewcommand\arraystretch{1.6}
  \caption{\label{tab:ope1/2}The coefficients $C_{ij}$ of the OPE potential for $J^P=\frac{1}{2}^-$ channels.}
  \begin{ruledtabular}
  \begin{tabular}{cccccc}
      $1 / 2^{-}$ & $\Lambda_{c} D$ & $\Lambda_{c} \bar{D}^{*}$ & $\Sigma_{c} \bar{D}$ & $\Sigma_{c} \bar{D}^{*}$ & $\Sigma_{c}^{*} \bar{D}^{*}$ \\
      \hline
      $\Lambda_{c} \bar{D}$ & 0 & 0 & 0 & $\frac{\sqrt{6}}{12} igg_2$ & $\frac{\sqrt{3}}{6} igg_2$ \\
      $\Lambda_{c} \bar{D}^{*}$ & & 0 & $-\frac{\sqrt{6}}{12} igg_2$ & $-\frac{\sqrt{2}}{6} gg_2$ & $\frac{1}{6} gg_2$ \\
      $\Sigma_{c} \bar{D}$ & & & 0 & $\frac{\sqrt{3}}{9} igg_1$ & $-\frac{\sqrt{6}}{18} igg_1$ \\
      $\Sigma_{c} \bar{D}^{*}$ & & & & $-\frac{2}{9} gg_1$ & $-\frac{\sqrt{2}}{18} gg_1$ \\
      $\Sigma_{c}^{*} \bar{D}^{*}$ & & & & & $-\frac{5}{18} gg_1$ \\
    \end{tabular}
    \end{ruledtabular}
\end{table}

\begin{table}
  \renewcommand\arraystretch{1.6}
  \caption{\label{tab:ope3/2}The coefficients $C_{ij}$ of the OPE potential for $J^P=\frac{3}{2}^-$ channels.}
  \begin{ruledtabular}
  \begin{tabular}{ccccc}
    $3 / 2^{-}$ & $\Lambda_{c} \bar{D}^{*}$ & $\Sigma_{c}^{*} \bar{D}$ & $\Sigma_{c} \bar{D}^{*}$ & $\Sigma_{c}^{*} \bar{D}^{*}$ \\
    \hline
    $\Lambda_{c} \bar{D}^{*}$ & 0 & $-\frac{\sqrt{6}}{12} gg_2$ & $\frac{\sqrt{2}}{12}gg_2$ & $\frac{\sqrt{10}}{12} gg_2$ \\
    $\Sigma_{c}^{*} \bar{D}$ & & 0 & $\frac{\sqrt{3}}{18}igg_1$ & $\frac{\sqrt{15}}{18} igg_1$ \\
    $\Sigma_{c} \bar{D}^{*}$ & & & $\frac{1}{9} gg_1$ & $-\frac{\sqrt{5}}{18} gg_1$ \\
    $\Sigma_{c}^{*} \bar{D}^{*}$ & & &  & $-\frac{1}{9} gg_1$\\
    \end{tabular}
    \end{ruledtabular}
\end{table}

\section{Numerical results and discussions\label{sec:numeric}}
\subsection{Fitting the LECs}

In this work, the $P_c(4312)$ is assigned to a $\frac{1}{2}^-$
$\Sigma_c\bar{D}$ state. The $P_c(4440)$ and $P_c(4457)$ are
assigned to the $\frac{1}{2}^-$ and $\frac{3}{2}^-$
$\Sigma_c\bar{D}^*$ states. As shown in Table~\ref{tab:ope1/2}, the
OPE potential is attractive in $\frac{1}{2}^-$ channel but repulsive
in $\frac{3}{2}^-$ channel. So we prefer to assign the $P_c(4440)$
to the $\frac{1}{2}^-$ channel. The fit under the opposite
assignment is shown in Appendix~\ref{appendix-ass}.

In our fits, the cutoff $\Lambda$ is fixed to 500 MeV, and there are
four LECs to be determined. However, the LEC $A$ only appears in the
diagonal terms of the $\Lambda_c\bar{D}^{(*)}$ channel. It weakly
affects the masses and widths through the coupled-channel effects.
Thus the fit is not sensitive to $A$. On the other hand, $A$ is
important to determine whether the $\Lambda_c\bar{D}^{(*)}$ systems
are bound. Thus we adopt two fitting strategies: 1) setting $A=0$;
2) letting $A$ varies to find the best fit.

The statistical uncertainties in the tables are estimated by the
condition $\chi^2\leq \frac{1+d.o.f.}{d.o.f.}\chi_0^2$, where
$\chi_0^2$ stands for the minimum of $\chi^2$. The pole positions in
Table~\ref{tab:fit1} are derived from the optimal set of LECs.

Table~\ref{tab:fit1} shows the results when we assign $P_c(4440)$ to
$\frac{1}{2}^-$ and $P_c(4457)$ to $\frac{3}{2}^-$. In this case,
$C_a$ plays a major role while the other LECs are relatively small.
The $C_a$ term provides an attractive central potential to bind the
$\Sigma_c^{(*)}$ and $\bar{D}$, while the OPE potential provides the
coupled-channel interactions and introduces the spin splitting
between the $P_c(4440)$ and $P_c(4457)$. Since the OPE potential in
$J^P=\frac{3}{2}^-$ $\Sigma_c\bar{D}^*$ system is repulsive, its
mass is larger. A similar relationship shows up in the
$\Sigma_c^*\bar{D}^*$ system. There are bound states in both
$\frac{3}{2}^-$ and $\frac{1}{2}^-$ channels, and the energy of the
$\frac{3}{2}^-$ state is higher. Their mass splitting is of the same
order of magnitude as the mass splitting between the $P_c(4440)$ and
$P_c(4457)$.

\begin{table}
  \renewcommand\arraystretch{1.3}
\caption{\label{tab:fit1}The fitting result when assigning
$P_c(4440)$ to $\frac{1}{2}^-$ and $P_c(4457)$ to $\frac{3}{2}^-$.
The units for LECs are GeV$^{-2}$, and the units for the pole
positions ($M-\frac{i\Gamma}{2}$) and cutoff $\Lambda$ are MeV. The
quantum numbers and main components are listed in parentheses.}
\begin{ruledtabular}
  \begin{tabular}{ccc}
    & Fit 1 & Fit 2  \\
    \hline
    $\chi^2/d.o.f$&1.13&1.08\\
    $\Lambda$&500&500\\
    $A$ & 0&   $-32_{-50}^{+15}$  \\
    $B$ & $2.3_{-3.8}^{+5.6}$& $-0.3_{-5}^{+5}$\\
    $C_a$ & $-53.0_{-3.0}^{+3.0}$&  $-59.0_{-6}^{+8}$\\
    $C_b$ & $1.3_{-4.8}^{+2.2}$& $4.3_{-10}^{+3}$\\
    $P_c(4312)$ & $4309.4-3.8i$($\Sigma_c\bar{D},\frac{1}{2}^-$)&$4312.4-5.1i$($\Sigma_c\bar{D},\frac{1}{2}^-$)\\
    $P_c(4440)$ & $4443.4-1.6i$($\Sigma_c\bar{D}^*,\frac{1}{2}^-$)&$4438.7-1.8i$($\Sigma_c\bar{D}^*,\frac{1}{2}^-$)\\
    $P_c(4457)$ & $4458.6-0.5i$($\Sigma_c\bar{D}^*,\frac{3}{2}^-$)&$4457.6-0.9i$($\Sigma_c\bar{D}^*,\frac{3}{2}^-$)\\
    other states & \begin{tabular}{c}
      $4377.8-1.6i$($\Sigma_c^*\bar{D},\frac{3}{2}^-$)\\
      $4503.9-0.5i$($\Sigma_c^*\bar{D}^*,\frac{1}{2}^-$)\\
      $4516.0-1.6i$($\Sigma_c^*\bar{D}^*,\frac{3}{2}^-$)
    \end{tabular}
    &\begin{tabular}{c}
      $4158.1-0.3i$($\Lambda_c\bar{D},\frac{1}{2}^-$)\\
      $4288.4-0.8i$($\Lambda_c\bar{D}^*,\frac{1}{2}^-$)\\
      $4292.6-1.7i$($\Lambda_c\bar{D}^*,\frac{3}{2}^-$)\\
      $4375.4-1.8i$($\Sigma_c^*\bar{D},\frac{3}{2}^-$)\\
      $4497.2-0.9i$($\Sigma_c^*\bar{D}^*,\frac{1}{2}^-$)\\
      $4513.0-2.6i$($\Sigma_c^*\bar{D}^*,\frac{3}{2}^-$)
    \end{tabular}
  \end{tabular}
\end{ruledtabular}
\end{table}

In Fit 2, the best fit reveals a large negative $A$. This results in
a bound $\Lambda_c\bar{D}^{(*)}$ state, while the influence on the
observed $P_c$ states is small. The widths of the
$\Lambda_c\bar{D}^{(*)}$ molecules arise from the imaginary part of
the OPE potentials, which is probably overestimated in our
approximation. Comparing Fit 1 and Fit 2, we conclude that the
coupled-channel interactions from OPE between
$\Lambda_c\bar{D}^{(*)}$ and $\Sigma_c\bar{D}^{(*)}$ are not
sufficient to generate a $\Lambda_c\bar{D}^{(*)}$ bound state.
Nevertheless, if the interaction in $\Lambda_c\bar{D}^{(*)}$ system
is attractive enough, there may exist three additional narrow
states.

The total width of $P_c(4312)$ is the largest, which physically
arises from the strong coupling and proximity of the
$\Lambda_c\bar{D}^*$ and $\Sigma_c\bar{D}$ channels. This differs
from the center value of the experimental result. One reason is the
uncertainties for the experimental widths are large, especially for
$P_c(4440)$. This reduces the weight of the width of $P_c(4440)$ in
the fit. Another important reason is that the width of $\Sigma_c$ is
not included in the calculation, which may influence the width of
the $P_c$ state by 1 to 2 MeV. For the states with $\Sigma_c^*$, the
width may increase by 10 MeV.

In contrast, the width of the $\frac{1}{2}^-$ $\Sigma_c\bar{D}^*$
bound state is small (except for the effect of the width of
$\Sigma_c^{(*)}$), although it is strongly coupled to the
$\frac{1}{2}^-$ $\Lambda_c\bar{D}$ channel and the phase space is
large. It indicates the influence from a relatively far threshold is
small. Since the momentum in the lower channel could be large and
the potential could be suppressed by the regulator, it is not quite
reasonable to consider the far-away thresholds in a non-relativistic
framework.

In all fits, we find the $J^P=\frac{3}{2}^-$ states near the
$\Sigma_c^*\bar{D}$ and the $\Sigma_c^*\bar{D}^*$ thresholds always
exist. They are on the 1st Riemann sheet with respect to the
$\Sigma_c^*\bar{D}^{(*)}$ threshold. The former corresponds to the
previously reported $P_c(4380)$ and the latter is predicted to be
located in the vicinity of 4520 MeV, whose width could be much
larger than our results since $\Sigma_c^*$ has a large width.

\subsection{Partial width}

The interaction between $\Lambda_c$ and $\bar{D}^{(*)}$ is believed
to be weak because the spin and the isospin of the light quarks in
$\Lambda_c$ are both zero. Thus we choose Fit 1 to calcultate the
branching ratios, root-mean-square (r.m.s.) radii and component
proportions of the $P_c$ states. The uncertainties are estimated in
the domain $\chi^2\leq \frac{1+d.o.f.}{d.o.f.}\chi_0^2$. The
uncertainties of $P_c(4312)$, $P_c(4440)$ and $P_c(4457)$ are
similar to the experimental uncertainties due to the fit, and the
uncertainties of widths are large.

The r.m.s. radius and the component proportion are defined by the
c-product \cite{Homma-cprod}
\begin{eqnarray}
  (\psi|r^2|\psi)=\sum_i\int r^2\psi_i(\bm{r})^2d^3\bm{r},\nonumber\\
  (\psi_i|\psi_i)=\int \psi_i(\bm{r})^2d^3\bm{r}\label{eq:c-prod},
\end{eqnarray}
in which the wave function of the $i$-th channel $\psi_i$ satisfies
the normalization condition
\begin{eqnarray}
  \sum_i(\psi_i|\psi_i)=1.
\end{eqnarray}
In the c-product, the inner product is defined using the square of
the wave function rather than the square of its modulus. Although
$\psi_i(r)$ is divergent at infinity in open channels, the integral
in Eq.~(\ref{eq:c-prod}) can be defined using analytical extension,
and is generally not real.

As shown in Table \ref{tab:expect}, the imaginary part of the r.m.s.
radius is small, which indicates the state is similar to the stable
states. The r.m.s. radii are of the order of magnitude of 1 fm and
qualitatively in proportion to the inverse of the binding energies,
which is in accordance with the molecular state assumption.

All the states are found to be the quasibound states. In other
words, the momenta with respect to the higher thresholds have
positive imaginary parts, which results in convergent wave functions
in coordinate space, while it is opposite for the lower thresholds.
In the higher channels, the pole is on the 1st Riemann sheet and the
divergent term of Eq.~(\ref{eq:fourier}) vanishes. The wave function
is similar to that of the bound states, which implies the
possibility of finding two free particles in infinity is zero.
Although the residue of the $T$ matrix in Eq.~(\ref{eq:Smatrix}) may
be large, the state will not decay into the corresponding channel.
Thus, when we evaluate the branching ratios, only the lower
channels, of which the pole lies on the 2nd Riemann sheet, are
considered. The three-body decays are partly included in the total
width, but not considered in the branching ratios.

Although the branching ratios in different channels are mostly
comparable, the $P_c(4312)$ is an exception. The state decays mainly
into the $\Lambda_c\bar{D}^*$ channel, while the decay to the
$\Lambda_c\bar{D}$ channel is suppressed since the OPE potential
vanishes in the $\Lambda_c\bar{D}^*\rightarrow\Lambda_c\bar{D}$ or
$\Sigma_c\bar{D}\rightarrow\Lambda_c\bar{D}$. In disregard of the
uncertainties, the branching ratios to the closer channels are
likely to be larger. For the $P_c(4457)$, the branching ratio to the
$\Sigma_c^*\bar{D}$ channel is larger. For the $P_c(4440)$ and
$P_c(4504)$, the branching ratios to the $\Lambda_c\bar{D}^*$
channel are relatively small.

We further calculate the proportion of components using the
probability defined by c-product and the results are listed in
Table~\ref{tab:wave functionratio}. The $P_c(4312)$ shows a mixing
of $\Lambda_c\bar{D}^*$ and $\Sigma_c\bar{D}$, while the other
states are nearly a single-channel bound state. Since they are bound
states in the corresponding channels, they can only decay to the
lower channels. The $P_c(4380)$ is a molecule of the
$\Sigma_c^*\bar{D}$. The $P_c(4440)$ and $P_c(4457)$ are the
molecules of $\Sigma_c\bar{D}^*$. The $P_c(4504)$ and $P_c(4516)$
are the molecules of $\Sigma_c^*\bar{D}^*$. The branching ratios are
derived from the probabilities.

\begin{table}
  \renewcommand\arraystretch{1.4}
\caption{\label{tab:expect}The root-mean-square radii and open-charm
branching ratios of the $P_c$ states. LECs in Fit 1 are adopted. The
unit for $M$ and $\Gamma$ is MeV, the unit for RMS radii is fm, and
the unit for the branching ratios is \%. "--" means the state will
not decay to the channel.}
\begin{ruledtabular}
  \begin{tabular}{cccc}
    $\frac{1}{2}^-$&$P_c(4312)$ & $P_c(4440)$ & $P_c(4504)$\\
    \hline
    $M$& $4309.4_{-2.5}^{+2.7}$ & $4443.5_{-3.5}^{+3.7}$ & $4504.0_{-4.7}^{+6.1}$ \\
    $\Gamma$& $7.8_{-6.6}^{+6.6}$ & $3.1_{-1.4}^{+0.8}$    & $1.5_{-1.4}^{+0.4}$ \\
    $M_{\text{exp}}$& $4311.9\pm 0.7^{+6.8}_{-0.6}$ & $4440.3\pm 1.3^{+4.1}_{-4.7}$ & \\
    $\Gamma_{\text{exp}}$& $9.8\pm2.7^{+3.7}_{-4.5}$ & $20.6\pm4.9^{+8.7}_{-10.1}$ & \\
    $\sqrt{(\phi_i|r^2|\phi_i)}$& $0.63-0.11i_{-0.07-0.09i}^{+0.07+0.09i}$ & $0.60-0.01i_{-0.01-0.00i}^{+0.03+0.01i}$  & $0.58+0.00i_{-0.01i-0.01i}^{+0.03+0.00i}$   \\
    $\Lambda_c\bar{D}$& $0.04_{-0.02}^{+0.01}$ & $10.8_{-2.7}^{+8.0}$ &  $8.7_{-6.6}^{+7.0}$   \\
    $\Lambda_c\bar{D}^*$& $99.96_{-0.01}^{+0.02}$ & $38.4_{-30.6}^{+24.9}$ &  $24.6_{-18.3}^{+17.1}$   \\
    $\Sigma_c\bar{D}$&--& $50.9_{-27.4}^{+38.6}$ &  $31.6_{-14.4}^{+16.2}$  \\
    $\Sigma_c\bar{D}^*$&--& -- & $35.2_{-9.7}^{+8.7}$    \\
    $\Sigma_c^*\bar{D}^*$&--& -- & --  \\
\hline
    $\frac{3}{2}^-$& $P_c(4380)$ & $P_c(4457)$ &$P_c(4516)$\\
    \hline
    $M$& $4377.9_{-3.0}^{+2.3}$ & $4458.6_{-2.5}^{+1.4}$ & $4516.0_{-2.5}^{+2.1}$\\
    $\Gamma$& $3.2_{-3.1}^{+1.7}$ & $1.0_{-0.4}^{+0.3}$ & $3.2_{-1.7}^{+1.4}$\\
    $M_{\text{exp}}$&  &$4457.3\pm 0.6^{+4.1}_{-1.7}$&\\
    $\Gamma_{\text{exp}}$& &$6.4\pm2.0^{+5.7}_{-1.9}$&\\
    $\sqrt{(\phi_i|r^2|\phi_i)}$& $0.74-0.03i_{-0.06-0.02i}^{+0.06+0.02i}$ & $0.84+0.01i_{-0.08-0.01i}^{+0.08+0.01i}$ & $0.67-0.01i_{-0.02-0.01i}^{+0.03+0.01i}$\\
    $\Lambda_c\bar{D}^*$& $100$ & $26.9_{-22.5}^{+30.0}$ & $18.1_{-14.5}^{+23.7}$\\
    $\Sigma_c^*\bar{D}$& -- & $73.1_{-30.0}^{+22.5}$ &  $45.6_{-13.1}^{+7.9}$\\
    $\Sigma_c\bar{D}^*$& -- & -- &  $36.2_{-10.6}^{+6.6}$\\
    $\Sigma_c^*\bar{D}^*$& -- & -- &  --\\

  \end{tabular}
\end{ruledtabular}
\end{table}

\begin{table}
  \renewcommand\arraystretch{1.4}
\caption{\label{tab:wave functionratio} The components of the $P_c$
states. The probability in the $i$-th channel is defined by the
c-product $(\phi_i|\phi_i)$ and hence has an imaginary part. The
unit is \%. }
\begin{ruledtabular}
  \begin{tabular}{cccc}
    $\frac{1}{2}^-$&$P_c(4312)$ & $P_c(4440)$ & $P_c(4504)$\\
    \hline
    $\Lambda_c\bar{D}$& $0.00_{-0.00}^{+0.00}$ & $0.07+0.01i_{-0.06-0.00i}^{+0.03+0.00i}$ &  $0.06+0.07i_{-0.06-0.04i}^{+0.03+0.02i}$   \\
    $\Lambda_c\bar{D}^*$& $\bm{5.3-0.5i}_{-4.6-0.9i}^{+5.5+3.9i}$ & $0.14-0.01i_{-0.12-0.01i}^{+0.10+0.01i}$ &  $0.08+0.05i_{-0.07-0.02i}^{+0.05+0.00i}$   \\
    $\Sigma_c\bar{D}$& $\bm{94.5+0.2i}_{-5.5-4.1i}^{+4.6+0.9i}$ & $0.11+0.04i_{-0.08-0.03i}^{+0.08+0.02i}$ &  $0.07-0.02i_{-0.05-0.01i}^{+0.06+0.02i}$  \\
    $\Sigma_c\bar{D}^*$& $0.2+0.3i_{-0.2-0.3i}^{+0.1+0.2i}$ & $\bm{99.1+0.2i}_{-0.4-0.1i}^{+0.5+0.1i}$ & $0.09+0.22i_{-0.07-0.16i}^{+0.04+0.18i}$    \\
    $\Sigma_c^*\bar{D}^*$& $0.02+0.05i_{-0.04-0.04i}^{+0.03+0.05i}$ & $0.6-0.2i_{-0.4-0.1i}^{+0.4+0.1i}$ & $\bm{99.7-0.3i}_{-0.1-0.2i}^{+0.2+0.2i}$   \\
    \hline
    $\frac{3}{2}^-$&$P_c(4380)$ & $P_c(4457)$&$P_c(4516)$\\
    \hline
    $\Lambda_c\bar{D}^*$& $0.24+0.17i_{-0.21-0.06i}^{+0.12+0.08i}$ & $0.01+0.00i_{-0.01-0.00i}^{+0.01+0.01i}$ & $0.03+0.08i_{-0.04-0.05i}^{+0.05+0.02i}$\\
    $\Sigma_c^*\bar{D}$& $\bm{99.6-0.3i}_{-0.3-0.1i}^{+0.2+0.2i}$ & $0.01+0.01i_{-0.01-0.01i}^{+0.01+0.02i}$ &  $0.08+0.03i_{-0.05-0.03i}^{+0.05+0.03i}$  \\
    $\Sigma_c\bar{D}^*$& $0.08+0.06i_{-0.08-0.05i}^{+0.13+0.03i}$ & $\bm{99.92-0.07i}_{-0.09-0.06i}^{+0.04+0.04i}$ &  $0.07+0.09i_{-0.05-0.07i}^{+0.05+0.08i}$\\
    $\Sigma_c^*\bar{D}^*$& $0.10+0.07i_{-0.10-0.06i}^{+0.19+0.04i}$ & $0.06+0.05i_{-0.03-0.03i}^{+0.08+0.05i}$ &  $\bm{99.81-0.19i}_{-0.06-0.09i}^{+0.09+0.09i}$\\

  \end{tabular}
\end{ruledtabular}
\end{table}

\section{Summary\label{sec:sum}}

We perform a deduction of the analytical extension of wave
functions in momentum space. Then the analytical behavior of the
wave function in coordinate space is obtained using the Fourier
transformation. We show how CSM works from the point of view of
analytical extension. Whether we include the residue of the pole of
the wave function in the integral or not will affect which Riemann
sheet the pole is located on. In this way, the branching ratio is
derived from the complex wave function. Such a formalism can be
easily extended to other systems.

In order to make use of the experimental values of the widths of the
$P_c$ states, we have performed a coupled-channel analysis using
CSM. The potential arises from OPE involving the on-shell three-body
intermediate states and contact terms with undetermined LECs. We use
the masses and widths of the $P_c(4312)$, $P_c(4440)$, $P_c(4457)$
as inputs to fit the LECs. Then we calculate the branching ratios of
the open-charm two-body final states of the observed $P_c$ states
and other predicted states.

Assuming the coupled-channel effects arise mainly from OPE, which
implies the LECs $B$ and $C_b$ are small, we prefer to assign the
$P_c(4440)$ to $\frac{1}{2}^-$ and the $P_c(4457)$ to
$\frac{3}{2}^-$. Under this assignment, three additional states are
obtained at the vicinity of 4380 MeV, 4504 MeV and 4516 MeV, which
are mainly the bound states of $\Sigma_c^*\bar{D}^{(*)}$. The mass
splitting of the latter two states is similar to that of the
$P_c(4440)$ and $P_c(4457)$, whereas their widths may be larger than
our prediction because of the large width of the $\Sigma_c^*$. If we
interchange the assignment, the $\frac{1}{2}^-$
$\Sigma_c^*\bar{D}^{*}$ may not be bound. Since the observed $P_c$
states depend weakly on the LEC $A$, its value is unlikely to be
determined. However, a large negative $A$ in the best fit will
result in extra states which are mainly $\Lambda_c\bar{D}^*$ bound
states. Given that the interactions in the $\Lambda_c\bar{D}$ system
are weak in meson-exchange models, we force $A$ to be zero and
calculate the branching ratios.

Our result shows that all the states are the quasibound states near
the physical region. The $P_c(4312)$ has considerable proportions in
the $\Lambda_c\bar{D}^*$ and $\Sigma_c\bar{D}$ channels. It lies on
the 1st Riemann sheet with respect to the $\Sigma_c\bar{D}$
threshold and the 2nd Riemann sheet with respect to the
$\Lambda_c\bar{D}^*$ threshold. It decays mainly to
$\Lambda_c\bar{D}^*$ rather than $\Lambda_c\bar{D}$. Other states
are mainly the bound states of the closest channel, and decay only
to the lower channels. The branching ratios of decaying to the
closer channels tend to be larger. These channels will be helpful to
search for the $P_c$ states.

\begin{acknowledgments}
{This work is supported by the National Science Foundation of China
under Grants No. 11975033, No. 12070131001 and No. 12147168. The
authors thank Yan-Ke Chen and Liang-Zhen Wen for helpful
discussions.}
\end{acknowledgments}

\appendix

\section{Interchanging the spin assignments\label{appendix-ass}}

Table~\ref{tab:fit2} shows the result when we assign $P_c(4440)$ to
$\frac{3}{2}^-$ and $P_c(4457)$ to $\frac{1}{2}^-$. In this case,
$C_b$ becomes important because it reverses the spin splitting
between the $\frac{3}{2}^-$ and $\frac{1}{2}^-$ states. One
remarkable feature is that $C_b$ introduces a large repulsive
potential in the $\frac{1}{2}^-$ $\Sigma_c^*\bar{D}^*$ channel, and
they are not bound anymore. There will be only one state around the
$\Sigma_c^*\bar{D}^*$ threshold with $J^P=\frac{3}{2}^-$. However,
if we allow $A$ to vary, there will be the $\Lambda_c\bar{D}^*$
bound states, and their mass splitting will not be reversed since
$C_b$ is in the $\Sigma_c^{(*)}$ sector.

\begin{table}
  \renewcommand\arraystretch{1.3}
\caption{\label{tab:fit2}The fitting result when assigning
$P_c(4440)$ to $\frac{3}{2}^-$ and $P_c(4457)$ to $\frac{1}{2}^-$.
The $\frac{1}{2}^-$ $\Sigma_c^*\bar{D}^*$ system is not bound. The
units for LECs are GeV$^{-2}$, and the units for the pole positions
($M-\frac{i\Gamma}{2}$) are MeV. The quantum numbers and main
components are listed in parentheses.}
 \begin{ruledtabular}
   \begin{tabular}{ccc}
     & Fit 3 & Fit 4  \\
     \hline
     $\chi^2/d.o.f$&1.58&0.92\\
     $\Lambda$&500&500\\
     $A$ & 0      &   $-38.3_{-20}^{+15}$  \\
     $B$ & $-0.1_{-2.8}^{+6.1}$& $-8.8_{-4.1}^{+5.4}$\\
     $C_a$ & $-55.4_{-3.9}^{+4.7}$&  $-67.1_{-4.3}^{+5.0}$\\
     $C_b$ & $-30.2_{-4.7}^{+5.0}$& $-28.3_{-3.8}^{+5.2}$\\
     $P_c(4312)$ &$4308.2 - 3.5i$($\Sigma_c\bar{D},\frac{1}{2}^-$) &$4311.9-4.9i$($\Sigma_c\bar{D},\frac{1}{2}^-$)\\
     $P_c(4440)$ &$4446.7 - 0.5i$($\Sigma_c\bar{D}^*,\frac{3}{2}^-$) &$4439.1-0.8i$($\Sigma_c\bar{D}^*,\frac{3}{2}^-$)\\
     $P_c(4457)$ &$4458.4 - 1.8i$($\Sigma_c\bar{D}^*,\frac{1}{2}^-$) &$4457.4-3.6i$($\Sigma_c\bar{D}^*,\frac{1}{2}^-$)\\
     other states & \begin{tabular}{c}
      $4377.5 - 1.6i$($\Sigma_c^*\bar{D},\frac{3}{2}^-$)\\
      $4526.7 - 0.2i$($\Sigma_c^*\bar{D}^*,\frac{3}{2}^-$)
    \end{tabular}
     &\begin{tabular}{c}
      $4154.2-0.6i$($\Lambda_c\bar{D},\frac{1}{2}^-$)\\
      $4277.1-0.8i$($\Lambda_c\bar{D}^*,\frac{1}{2}^-$)\\
      $4285.9-2.2i$($\Lambda_c\bar{D}^*,\frac{3}{2}^-$)\\
      $4372.7-1.8i$($\Sigma_c^*\bar{D},\frac{3}{2}^-$)\\
      $4524.5-1.4i$($\Sigma_c^*\bar{D}^*,\frac{3}{2}^-$)\\
      $4526.9-0.3i$($\Sigma_c^*\bar{D}^*,\frac{1}{2}^-$)
    \end{tabular}
   \end{tabular}
 \end{ruledtabular}
 \end{table}

\bibliography{Pc}

\end{document}